\documentstyle[psfig]{mn}
\newcommand{\beq}{\begin{equation}}
\newcommand{\eeq}{\end{equation}}
\newcommand{\bdm}{\begin{displaymath}}
\newcommand{\edm}{\end{displaymath}}
\newcommand{\bfig}{\begin{figure}}
\newcommand{\efig}{\end{figure}}
\newcommand{\msun}{M_{\odot}}
\def\etal{{\it et al.~}}
\def\ie{{\frenchspacing\it i.e. }}

\def\gtsima{$\; \buildrel > \over \sim \;$}
\def\ltsima{$\; \buildrel < \over \sim \;$}
\def\prosima{$\; \buildrel \propto \over \sim \;$}
\def\gsim{\lower.5ex\hbox{\gtsima}}
\def\lsim{\lower.5ex\hbox{\ltsima}}
\def\simgt{\lower.5ex\hbox{\gtsima}}
\def\simlt{\lower.5ex\hbox{\ltsima}}
\def\simpr{\lower.5ex\hbox{\prosima}}

\title[Dust Formation in Primordial Type II Supernovae]{Dust Formation in 
Primordial Type II Supernovae}

\author[P. Todini \&  A. Ferrara]{Paolo Todini$^1$ and  Andrea Ferrara$^{2,3}$\\
$^1$ Universit\`a degli Studi di Firenze, Dipartimento di
                    Astronomia, Largo Enrico Fermi 5, 50125 Firenze, Italy \\
$^2$ Osservatorio Astrofisico di Arcetri, Largo Enrico Fermi 5,
      50125 Firenze, Italy \\
$^3$ Center for Computational Physics, University of Tsukuba, Tsukuba-shi, Ibaraki-ken, 305-8577, Japan}

\date{August  2000}
\begin{document}
\maketitle

\begin{abstract}
We have investigated the formation of dust in the ejecta of Type II supernovae (SNe), 
mostly of   primordial composition, to answer the question of where are the first 
solid particles formed in the universe. However, we have also considered non-zero
progenitor's metallicity values up to $Z=Z_\odot$. The calculations are based on standard
nucleation theory and the scheme has been first tested on the well studied case
of SN1987A, yielding results that are in agreement with the available
data. We find that:  
{\it i)} the first dust grains are predominantly made of silicates,  
amorphous carbon (AC), magnetite, and corundum; 
{\it ii)} the largest grains are the AC ones, with sizes around 300\AA~, 
whereas other grain types have smaller radii, around 10-20\AA. The grain size 
distribution depends somewhat     on the thermodynamics of the ejecta expansion
and variations in the results by a factor $\approx 2$ might occur within   
reasonable estimates of the relevant parameters.
Also, and for the same reason, the grain size distribution, is essentially
unaffected by metallicity changes.
The predictions on the amount of dust formed are very robust:
for $Z=0$, we find that SNe with masses in the range (12-35)$M_\odot$
produce about $0.08 M_\odot \simlt M_d \simlt 0.3 M_\odot$ of dust/SN.
The above range increases by roughly 3 times as the metallicity
is increased to solar values.
We discuss the implications and the cosmological consequences of the results.
\end{abstract}

\section{Introduction}

Our understanding of galaxy formation is currently making tremendous advances
and recent investigations have also focused on the formation of the first luminous
sources (often referred to as PopIII objects). 
Several difficult questions arise when one deals with these peculiarly small 
collapsed objects (for a discussion see Ferrara 2000) primarily concerning the
properties of their first stars and  IMF (Tegmark \etal 1997, Abel \etal 1998, 
Bromm, Coppi \& Larson 2000,  Omukai \& Nishi 1999, 
Susa \& Umemura 2000, Nakamura \& Umemura 2000, Ripamonti \etal 2000), their response 
to the energy injection of supernovae (SN) (MacLow \& Ferrara 1999, 
Ciardi \etal 2000), 
their ability to form and preserve enough H$_2$ to provide the cooling for collapse
(Ciardi, Ferrara \& Abel 2000, Haiman, Abel \&  Rees  2000, Machacek, Bryan \& 
Abel 2000), and their contribution to the reionization (Gnedin \& Ostriker 1997,
Gnedin 2000, Ciardi \etal 2000, Ciardi \etal 2000a) and metal enrichment 
(Ferrara, Pettini \& Shchekinov 2000) of the IGM.

In  spite of this flourishing activity little attention has been given to the
role of dust in these early epochs. At lower redshifts the dramatic effects of dust
have been appreciated when estimates of the cosmic star formation rate (SFR) 
were attempted via UV/visible surveys of distant galaxies. It was soon 
realized that approaches based on the
``dropout" technique are poorly sensitive to dust-obscured galaxies.
Hence, the SFR deduced in this way could represent a severe
underestimate of the actual one, if even a rather modest amount of dust is
present in the interstellar medium of the star forming galaxy. Also, some
galaxies could be so heavily extinguished that they could be completely
missed from the UV/visible census (Cimatti \etal 1997, Ferrara \etal
1999).          

Direct indications of the existence
of dust at high redshift come from the reddening of
background quasars; indirect evidences of dust in damped Ly$\alpha$ systems 
have been obtained from the relative gas-phase abundances of Zn and Cr (Pettini
\etal 1997). Fall \etal (1996) have calculated the cosmic infrared background
from dust in damped Ly$\alpha$ systems, and found good agreement
with FIR background deduced from {\it COBE}/FIRAS data, 
which also seem to imply the presence of dust.
Recent detection of heavy elements, such as carbon and silicon (Lu \etal 1998,
Cowie \& Songaila 1998, Ellison \etal 2000)
in very low column density Ly$\alpha$ clouds ($\log N_{HI} < 14$) at redshift $z\sim
3$ can potentially indicate that dust exists also in the  Ly$\alpha$  forest: it is
quite natural to assume that dust grains are associated with  heavy elements.
Dust in the forest clouds would be relevant to the
understanding of their origin and association to
PopIII objects, the heavy element enrichment pattern of intergalactic medium, 
and the thermal history of Ly$\alpha$ clouds (Ricotti \& Gnedin 2000, Schaye 
\etal 2000). 

The questions that we pose here are the following. When was dust first formed ?
Is grain formation possible starting from a metal-free environment ? What 
are the dust properties and amount produced ? How are these quantities affected
by metallicity changes ? 

Dwek \& Scalo (1980) have shown that dust injection in the ISM of the Galaxy 
from supernovae dominates
other sources, if indeed grains can form and survive in the ejecta.
This has become clear after the SN1987A event, in which dust has been
unambiguously detected (Moseley \etal 1989, Kosaza, Hasegawa \&
Nomoto 1989).
Indeed, the bulk of the refractory elements (characterized by higher melting
temperatures, such as Si, Mg, Fe, Ca, Ti, Al etc.)
is injected into the ISM by supernovae (McKee 1989).
At high redshift, the contribution to dust production due to evolved stars
(M and carbon stars, Wolf-Rayet stars, red giants and supergiants, novae) 
is even more negligible or absent. The reason is that the typical evolutionary 
timescale of these stars ($\simgt 1$ Gyr) is longer than the age of the universe,
$t_H = 6.6 h^{-1}(1+z)^{-3/2}$~Gyr in a EdS cosmology, if  $(1+z) \simgt 5$ 
(adopting $h=0.65$). Thus,
it seems clear that if high redshift dust exists, it must have been produced 
by Type II SNe, to which we then devote the rest of this study.

\section{Dust Formation Model}
\subsection{Dust nucleation and accretion}
The formation of solid materials from the gas phase can occur only from a vapor in a supersaturated state. 
Because of the existence of a well defined
condensation barrier, expressed by a corresponding ``critical cluster'' size, the formation of solid 
particles in a gaseous medium is described
as a two-step process: {\it i)} the formation of critical clusters;
{\it ii)} the growth of these clusters into macroscopic dust grains.
The classical theory of nucleation (Feder \etal  1966, Abraham 1974) gives 
an expression for the 
nucleation current, $J$, \ie the number of clusters
of critical size formed per unit volume and unit time in the gas:
\beq
J=\alpha \Omega \bigg(\frac{2\sigma}{\pi m_1}\bigg)^{1/2}c_1^2 \exp \bigg\{-\frac{4\mu^3}{27(\ln S)^2} 
\bigg\},
\label{eq:J}
\eeq
where $\mu=4\pi a_o^2\sigma /k_BT$ with $a_o$ the radius of molecules (or atoms, depending on chemical 
species) in the condensed phase; $\sigma$ is
the specific surface energy (corresponding to surface tension in liquids), $k_B$ is the Boltzmann 
constant and $T$ the gas      temperature;
$m_1$ and $c_1$ are the mass and the concentration of the monomers in the gas phase,  respectively; 
$\Omega=(4/3)\pi a_o^3$ is the volume of the single
molecules in the condensed phase, $\alpha$ is the sticking coefficient and $S$ 
the supersaturation ratio, defined below. The subsequent growth of the clusters occurs by accretion and 
is described by:
\beq
\frac{dr}{dt}=\alpha \Omega v_1 c_1(t),
\label{eq:drdt}
\eeq
with the condition:
\beq
r(0)=r_* =\frac{2\sigma \Omega}{k_B TlnS},
\eeq
where $v_1$ is the mean velocity of monomers, $r(t)$ is the cluster radius
at time $t$, and 
$r_*$ is the cluster critical radius. Equations 
(\ref{eq:J}) and (\ref{eq:drdt}) describe nucleation and growth of solid particles in a gas composed 
of a single chemical species (\ie reactions of the type
$Fe(gas) \rightarrow Fe(solid)$ or $SiO(g) \rightarrow SiO(s)$). However,
there are some compounds (like 
forsterite, $Mg_2SiO_4$) whose nominal molecule
does not exist in the gas phase. These compounds form directly in the solid phase by means of a 
chemical reaction with the reactants in the gas 
phase. We need to extend the theory described above to this situation. Following Kozasa \& Hasegawa 
(1987, see also Hasegawa \& Kozasa 1988) we
consider a vapor in a supersaturated state. In this vapor grains condense homologously via the reaction:
\beq
\sum_i \nu_i A_i= \textrm{solid compound},
\label{eq:reaction}
\eeq
where $A_i$'s represent the chemical species of reactants and products in the gas phase and $\nu_i$'s 
are stoichiometric coefficients, which are 
positive for reactants and negative for products respectively. We make the following assumptions:
{\it i)} the rates of nucleation and grain growth are controlled by a
single chemical species, referred to as a key species.
{\it ii)} the key species corresponds to the reactant with the least 
collisional frequency onto a target cluster.
In this case, eqs. (\ref{eq:J}) and (\ref{eq:drdt}) become:
\beq
J=\alpha \Omega \bigg(\frac{2\sigma}{\pi m_{1k}}\bigg)^{1/2}c_{1k}^2 \exp \bigg\{-\frac{4\mu^3}
{27(\ln S)^2}\bigg\},
\eeq
and
\beq
\frac{dr}{dt}=\alpha \Omega v_{1k} c_{1k}(t),
\eeq
where $m_{1k}, c_{1k}$ and $v_{1k}$ are the mass, concentration and mean velocity of 
monomers of key species, respectively. 
In this case the supersaturation
ratio is expressed by:
\beq
\ln S= -\frac{\Delta G_r}{RT} +\sum_i \nu_i \ln P_i,
\label{eq:lns}
\eeq
where $P_i$ is the partial pressure of the i-th specie, $R$ is the gas constant and $\Delta G_r$ is 
the Gibbs free energy 
for the reaction (\ref{eq:reaction}).\\
We investigate the formation of the following solid compounds: $Al_2O_3$ (corundum), iron, $Fe_3O_4$ 
(magnetite), $MgSiO_3$ 
(enstatite), $Mg_2SiO4$ 
(forsterite) and amorphous carbon grains (ACG). These compounds are constituted by the most abundant heavy elements in the ejecta. 
Numerical constants used in our
calculations are summarized in Tab. 1. The value of the sticking coefficient $\alpha$ is set equal 
to 1 for all reactions; we have checked that the final results are
insensitive to a different choice in the plausible range 
$\alpha = 0.01 \div 1$). 
\begin{table*}
\centering
\begin{minipage}{120mm}
\caption{Chemical reactions and numerical constants used in dust formation calculations.
          $^a$Kozasa, Hasegawa \& Nomoto 1989; $^b$Kozasa \etal  1996; $^c$Hasegawa \& Kozasa 1988.}
\begin{tabular}{|l|c|c|c|} 
\hline 
solid compound & chemical reaction & $\sigma$ [erg][cm]$^{-2}$ & $a_o$ [$10^{-8}$ cm] \\ 
\hline
ACG & $C(g) \rightarrow C(s)$ & $1400^a$ & 1.28 \\ 
$Al_2O_3$ & $2Al+3O \rightarrow Al_2O_3$ & $690^a$ & 1.72  \\
$Fe$ & $Fe(g) \rightarrow Fe(s)$ & $1800^c$ & 1.41 \\
$Fe_3O_4$ & $3Fe+4O \rightarrow Fe_3O_4$ & $410^a$ & 1.80 \\
$MgSiO_3$ & $Mg+SiO+2O \rightarrow MgSiO_3$ & $400^b$ & 2.32 \\
$Mg_2SiO_4$ & $2Mg+SiO+3O \rightarrow Mg_2SiO_4$ & $436^a$ & 2.05 \\ 
\hline
\label{tabular:tab1}
\end{tabular}
\end{minipage}
\end{table*}

\subsection{Supernova model}
We now     describe the adopted model for the SN        ejecta. Before the
explosion the progenitor develops the standard "onion skin" stratified 
structure, with a hydrogen-rich envelope, a helium layer, and several 
thinner heavy element layers up to a  $Fe-Ni$ core. During the explosion a 
shock wave propagates through the layers, reheats the gas and triggers the 
explosive nucleosynthesis phase. This phase lasts for few hours,
then expansion cools the gas and the thermonuclear reactions turn off. After the explosion the SN        starts to expand homologously, with velocity 
$v \propto R$, where $R$ is the distance from the center.    
During the first weeks Rayleigh-Taylor instabilities cause the
mixing of the internal layers (Fryxell, M\"uller \& Arnett 1991). The early emergence of X-rays and $\gamma$-rays observed in SN 1987A (Itoh \etal 
1987; Kumagai \etal  1988) can be explained if radioactive $^{56}Co$ is mixed from the internal regions of the star into the external ones; more
precisely, observations suggest mixing of the materials in the ejecta at least up to the outer edge of the helium layer. Dust grains are formed by
heavy elements so we focus on          the volume that containing them, \ie the 
sphere of radius $R$, defined as the radius of the outer edge of the He-rich
layer. It is thought that mixing forms clumps of heavy elements embedded in the He-rich layer. As a first approximation, we assume that
mixing is complete, and that the gas has uniform density and temperature 
in the considered volume at any given      time.

Photometric observations have shown that a SN emits typically $10^{49}$ erg in electromagnetic energy, but current theoretical  models predict 
kinetic energies $E_{kin}\approx 10^{51}$ erg. 
The expansion velocity $v$ is then given by:
$v\simeq \sqrt{E_{kin}/M_{tot}}$,
where $M_{tot}$ is the total mass ejected  by the
SN. We take the chemical composition
of the expelled gas from the results of Woosley \& Weaver 1995 (hereafter WW95),
apart from the specific case of SN 1987A, see below. They determine the nucleosynthetic yields 
of isotopes lighter than A=66 (Zinc) for
a grid of stellar masses and metallicities including stars in the mass range 
$11-40~\msun$ and metallicities
$(Z/Z_\odot)=0, 10^{-4}, 0.01, 0.1, 1$.
They also give the values for $E_{kin}$ and $M_{tot}$ for all the SN 
models considered. The range $11-40~\msun$ is the most relevant
mass range for the production of  heavy elements. In fact, stars with
mass between 8 and 11 $\msun$ are characterized by very thin heavy
element layers, whereas stars heavier than 40 $\msun$ might be rare
and give rise to a black hole partially swallowing the nucleosynthetic
products (Maeder 1992). 

Expansion of the ejecta leads to cooling of the gas. We have already mentioned
that the radiation losses are only a few percent of the total internal 
energy; their contribution to cooling is even smaller in the first week 
after the explosion due to the high opacity which prevents photons from
escaping from the inner regions. Therefore, it is a good approximation
to assume that the expansion is adiabatic, although this hypothesis becomes 
less correct in the advanced evolutionary stages. 
Radiation losses are also    partly balanced by the heating provided by
radioactive decay (especially of $^{56}Ni \rightarrow ^{56}Co \rightarrow 
^{56}Fe$). We neglect here these complications and assume that  
the expansion is adiabatic. In this case (for a perfect gas) the
temperature evolution is given by 
\begin{displaymath}
T=T_i \bigg( 1+\frac{v}{R_i}t \bigg)^{3(1-\gamma)};
\end{displaymath}
$\gamma$ is the adiabatic index,    $T_i$ and $R_i$ are the temperature
and the radius at the beginning of the computation, $t$ is the time elapsed 
from this initial epoch. At the temperatures of 
interest ($T < 6000$ K) the gas density is $\approx 10^8$ atoms cm$^{-3}$ so the use of the perfect gas law is well justified. We set the
values of $R_i$, $T_i$ and $\gamma$ as follows. From photometric observations 
of SN 1987A (Catchpole \etal  1987) it is deduced that photosphere and the outer
edge of $He$-rich layer overlap $\approx 70$ days after the explosion, 
when the photospheric temperature is 5400 K and the radius $R= 1.6 
\times 10^{15}$ cm. For the adiabatic index  we take $\gamma = 1.25$ as in Kozasa, Hasegawa 
\& Nomoto (1989). We generally use these fiducial values in our models, but  
we will consider the effects of varying the values of $R_i$ and $\gamma$ 
when discussing the results.

\subsection{Molecule formation}
In the ejecta of SN 1987A molecules of $CO$ were detected for the first time in a SN 
(Meikle \etal  1989; Meikle \etal  1993; Bouchet
\& Danziger 1993) together with  $SiO$ (Aitken \etal  1988; Bouchet \etal  1991). These two molecules are very important for our study because carbon atoms bound
in $CO$ are not available to form ACG, whereas $SiO$ molecules take part 
in the chemical path leading to the formation of $MgSiO_3$ and $Mg_2SiO_4$. 
We investigate the process of molecule formation under the assumption
of chemical equilibrium. In the absence of grains, molecular formation in 
the gas phase must
have been initiated by radiative processes. We assume that formation of $CO$ is dominated by
radiative association (Lepp, Dalgarno \& McCray 1990;
Liu, Dalgarno \& Lepp 1992):
\bdm
C+O \longrightarrow CO+h\nu,
\edm
with a rate coefficient $K_{ra}(CO)$. The main destruction process of 
$CO$ is the impact with the energetic electrons produced by the  
radioactive decay of $^{56}Co$ (Liu \& Dalgarno 1995):
\bdm
CO+e \longrightarrow C+O+e,
\edm
with a rate coefficient $K_{rd}(CO)$. In steady state, the abundance of $CO$ is given by:
\beq
n(CO)=\frac{K_{ra}(CO)}{K_{rd}(CO)}n(C)n(O).
\eeq
Analogously, radiative association is the most important mechanism for
the formation  of $SiO$ (Liu \& Dalgarno 1996) via the reaction:
\bdm
Si+O \longrightarrow SiO+h\nu,
\edm
with rate coefficient $K_{ra}(SiO)$. Silicon monoxide is mainly destroyed by impact with energetic electrons:
\bdm
SiO+e \longrightarrow Si+O+e,
\edm
with rate coefficient $K_{rd}(SiO)$ and by charge transfer with positive ions of $Ne$ (Liu \& Dalgarno 1996):
\bdm
SiO+Ne^+ \longrightarrow Si^++O+Ne,
\edm
with rate coefficient $K_{ct}(SiO)$. The abundance of $SiO$ is given by:
\beq
n(SiO)=\frac{K_{ra}(SiO)}{K_{rd}(SiO)+K_{ct}(SiO)n(Ne^+)}n(Si)n(O).
\eeq
We assume for the rate coefficients the following values:
\begin{eqnarray}
K_{ra}(CO)&=&10^{-16}\times(-0.0398+1.25T_4-1.46T_4^2+ \nonumber \\
& &0.88T_4^3-0.21T_4^4)\quad \textrm{cm$^3$s$^{-1}$}, \nonumber
\end{eqnarray}
(Gearhart, Wheeler \& Swartz 1999, Dalgarno, Du \& You 1990), where $T_4=T/10^4$ K.
\bdm
K_{ra}(SiO)=5.52\times10^{-18} T^{0.31} \quad \textrm{cm$^3$s$^{-1}$},
\edm
(Andreazza \etal  1995, Liu \& Dalgarno 1996).
\bdm
K_{ct}(SiO)=2\times10^{-12} \quad \textrm{cm$^3$s$^{-1}$},
\edm
(Liu \& Dalgarno 1996). The ejecta are only moderately ionized with fractional ionization $\approx 10^{-2}$; hence, we assume $n(Ne^+)=0.02n(Ne)$. \\
To calculate the rate coefficient $K_{rd}(CO)$, we assume the same rate of destruction by energetic electron impact for $CO$ as for $SiO$, \ie 
$K_{rd}(CO)=K_{rd}(SiO)$. High energy X-rays and $\gamma$-rays produced by the chain of radioactive decay $^{56}_{28}Ni$ $\rightarrow$
$^{56}_{27}Co$ $\rightarrow$ $^{56}_{26}Fe$ interact by Compton scattering with the electrons in the ejecta. The average energy 
deposition rate per particle in the ejecta is (Woosley, Pinto \& Hartmann 1989):
\begin{eqnarray}
L_{\gamma}&=&7.5\times10^{-8}\frac{N_{i}(^{56}Co)}{N_{tot}}\langle E_{\gamma} \rangle f_{\gamma}(K_{56}) \nonumber \\
& &\exp \bigg\{-\frac{t}{\tau_{56}}\bigg\} \quad \textrm{MeV s$^{-1}$}. \nonumber
\end{eqnarray}
$N_{i}(^{56}Co)$ is the total number of atoms of $^{56}Co$ in the ejecta; $N_{tot}$ is the total number of particles in the gas;
$\langle E_{\gamma}\rangle =3.57$~MeV is the mean energy of $\gamma$-ray released by each decay; $\tau_{56}=111.26$~days is
the e-folding time of $^{56}Co$. 
The deposition function, $f_{\gamma}$ is proportional to the fraction of
trapped $\gamma$-photons,
\bdm
f_{\gamma}(K_{56})=1-\exp\{-K_{56}\phi_o(t_o/t)^2\},
\edm
with $\phi_o=\phi(t_o)$, the mass column density of the ejecta 
at some fiducial time $t_o$, and $K_{56}$ an average opacity for 
$^{56}Co$ decay $\gamma$-rays. We assume $\phi_o=7\times 10^4$ g cm$^{-2}$ at $t_o=10^6$ s and $K_{56}=0.033$ cm$^2$ g$^{-1}$. These 
values are appropriate for SN 1987A but, lacking more detailed
information, we extrapolate them to all our models. Finally, the
estimated destruction rate of $CO$ and $SiO$ by energetic electron
impact is 
\bdm
K_{rd}(CO)\equiv K_{rd}(SiO)=\frac{L_{\gamma}}{W_d} \quad \textrm{s$^{-1}$},
\edm
with $W_d$ being the mean energy per dissociation, defined as the energy of 
primary electrons divided by the number of molecule dissociations
(Liu \& Victor 1994). For a fractional ionization of the gas $\approx 
10^{-2}$, $W_d=152$~eV.

\section{A test case: SN 1987A}
To test and calibrate our dust formation model, we first apply it to 
SN 1987A, a case for which firm evidences of dust formation have been collected. 
In view of this test, we briefly summarize the relevant observational results. \\

\subsection{Observational results}
The most relevant evidence of newly formed dust grains in the ejecta of SN 1987A is the blue shift of the line profiles. Spectroscopic
observations detected this change between August 1988 and March 1989 
(Lucy \etal  1989, Lucy \etal  1991). This effect is likely to be
caused by the larger  attenuation suffered by radiation received from 
receding matter due to dust grains distributed in the ejecta. 
The condensation
efficiency, \ie the dust mass expressed as a fraction of the maximum value 
permitted by the elemental abundances in the ejecta, derived
by authors of observations from their data is $\leq 10^{-3}$. 
A further interesting observation is the stronger fading of the $[SiI]\lambda
1.65\mu$m line flux relative to the continuum after day 530. This 
can be interpreted as depletion of $Si$ from the gas phase as a
consequence of the formation of silicate grains. If the line fading is due solely to depletion, than the condensation efficiency rises to
$> 50$\%. 

Another evidence of the formation of dust is an  IR continuum excess   
over that expected from a Planck spectrum fitted to
the SN        emission at optical wavelengths (Roche \etal  1989, Wooden \etal. 1993).
Dust grains extinguish UV-visible radiation from the central energy source, 
re-emitting in the IR bands. 
This process might be responsible for the observed increase in the 10 
and 20$\mu$m fluxes around day 350$\div$450 (Roche \etal 
1989, Bouchet \& Danziger 1993, Meikle \etal  1993).

\subsection{Model Results}             
\begin{figure}
\psfig{figure=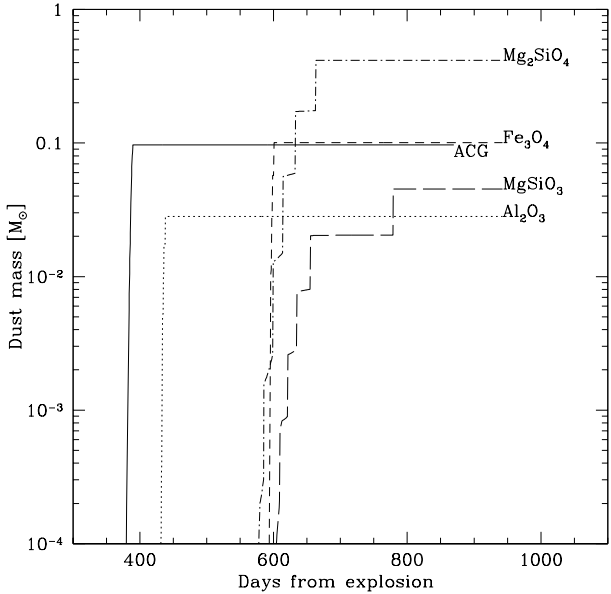,height=9cm}
\caption{Dust formation as a function of time elapsed since explosion in SN1987A}
\label{fig1}
\end{figure}

We now turn to the main results from our nucleation numerical computations
obtained by solving the above equations. The chemical composition of the SN
1987A ejecta is taken from Nomoto \etal 1991. The value of expansion velocity 
of gas is set to $v=2100$ km s$^{-1}$. This is the minimum expansion velocity 
determined from the
HI P$\alpha$ absorption trough (McGregor 1988, Nulsen \etal  1990) and is thought to represent the expansion velocity of the 
inner edge of the hydrogen envelope. 
Fig. \ref{fig1} shows the mass of dust formed in the ejecta as a function
of the time elapsed since explosion for the different solid compounds
found to be present.
We note that there are two episodes of dust formation. In the first one, ACG 
(formation time $t=380$~days) and $Al_2O_3$ ($t=430$~days) grains are formed.
This process is likely to be responsible for the 
IR-excess observed at that time.   In the second episode magnetite and
silicate grains are formed, at about $t=600$~days. Two points are worth
noticing to this concern. 
The first one is that this epoch corresponds to the formation of the
predominant fraction of dust mass of the SN (0.57 $M_\odot$
corresponding to about 84\% of the total amount); this dust formation
episode might be responsible for          
the blueshift of line profiles, observed only after day 530, when ACG and $Al_2O_3$ formed.

The second point is that the formation of $MgSiO_3$ and $Mg_2SiO_4$ might be 
related to the fading of the $Si$ and $Mg$ lines. The hypothesis
that these elements form enstatite and forsterite would also be suggested 
by the behavior of the $SiO$ molecule. 
\begin{figure}
\psfig{figure=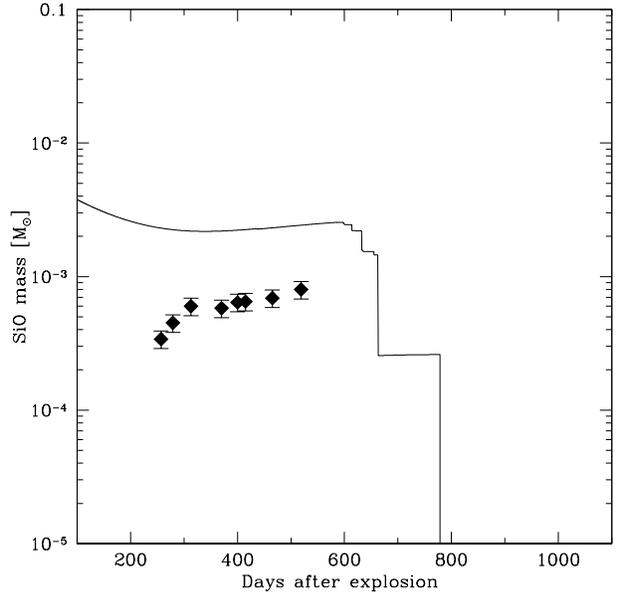,height=9cm}
\caption{Evolution of the $SiO$ mass (solid line) as a function of time compared with
the observational data (points) taken from Liu \& Dalgarno (1996).}
\label{fig2}
\end{figure}
The 
silicon monoxide rotovibrational line ($v = 1\rightarrow 0$, $\lambda = 7.8 \mu$m) emission
was detected after 160 days (Aitken \etal  1988) and it remained clearly
detectable until 519 days (Bouchet \etal  1991) but is no longer detected at 578 days (Roche \etal  1989). The time behavior of $SiO$ emission
can be understood by inspecting Fig. \ref{fig2}, where we show the predicted
$SiO$ mass vs. the observed mass as a function of time. Note the rapid fall 
at 660 days, 70 days after the beginning of silicate
formation, probably due to the depletion of $Si$ atoms; also our model
seems to slightly overpredict (by  a factor 3.1) the amount of $SiO$ produced at early
times. This could be due to our simplified treatment of the chemical
network for this molecule or inaccuracies in the rate coefficients. 
The dust formation efficiency deduced here           is 95\%, consistent 
with that deduced from the $Si$ line fading. 

All together, we look at the above results as a satisfactory success of
our model in reproducing, at least qualitatively,  the principal features 
of dust formation in SN 1987A.

\section{Dust in Primordial Supernovae}
In this Section we present the general results concerning dust
formation in primordial SNe.
We take the chemical compositions of the gas from Tables 16A and 16B of WW95, 
and the relevant results are reported in Tab. 2.
\begin{table*}
\centering
\begin{minipage}{180mm}
\caption{Adopted chemical composition of the supernova ejecta ($2.5\times 10^4$~s 
after explosion) for metallicity $Z=0$. Data are taken from WW95 and corrected to take into account
radioactive decay.}
\begin{tabular}{|l|r r r r r r r r r r}
\hline \hline
 & $12\msun$ & $13\msun$ & $15\msun$ & $18\msun$ & $20\msun$ &
$22\msun$ & $25A \msun$ & $30A\msun$ & $35A\msun$ & $40A\msun$ \\
 & & & & & & & $25B \msun$ & $30B\msun$ & $35B\msun$ & $40B\msun$ \\
\hline
$He$ & 4.08E+00 & 4.42E+00 & 4.90E+00 & 5.70E+00 & 6.32E+00 & 7.10E+00 & 7.33E+00 & 9.11E+00 & 8.17E+00 & 8.63E+00\\
 & & & & & & &7.82E+00 & 9.30E+00 & 1.06E+01 & 1.20E+01\\
$C$  & 4.30E--02 & 6.84E--02 & 1.45E--01 & 1.10E--01 & 8.98E--02 & 2.77E--01 & 8.31E--02 & 1.09E--01 & 1.25E--09 & 7.73E--10\\
 & & & & & & &4.23E--01 & 3.48E--01 & 3.49E--01 & 2.85E--01\\
$N$  & 5.41E--06 & 1.00E--05 & 2.46E--05 & 2.13E--06 & 1.76E--06 & 8.39E--05 & 2.39E--04 & 3.64E--05 & 2.13E--08 & 2.66E--08\\
 & & & & & & &3.29E--04 & 2.74E--03 & 6.29E--05 & 6.17E--06\\
$O$  & 6.67E--02 & 1.37E--01 & 4.00E--01 & 3.00E--02 & 9.21E--03 & 1.85E+00 & 9.67E--03 & 2.65E--02 & 3.08E--10 & 4.19E--10\\
 & & & & & & &2.33E+00 & 4.35E+00 & 1.92E+00 & 5.78E--01\\
$Ne$ & 2.61E--03 & 3.04E--02 & 9.28E--02 & 7.72E--04 & 2.12E--04 & 5.57E--01 & 5.79E--07 & 4.66E--06 & 1.37E--12 & 1.95E--12\\
 & & & & & & &5.10E--01 & 1.04E+00 & 3.76E--01 & 1.10E--02\\
$Na$ & 9.83E--06 & 1.90E--04 & 5.68E--04 & 9.57E--10 & 1.01E--09 & 3.90E--03 & 1.04E--08 & 5.04E--08 & 4.02E--13 & 6.37E--13\\
 & & & & & & &2.85E--03 & 3.88E--03 & 1.86E--03 & 1.07E--07\\
$Mg$ & 3.88E--03 & 1.50E--02 & 3.41E--02 & 4.80E--06 & 2.04E--06 & 9.52E--02 & 5.61E--08 & 5.90E--07 & 2.30E--13 & 3.33E--13\\
 & & & & & & &9.38E--02 & 2.49E--01 & 4.83E--02 & 2.89E--04\\
$Al$ & 7.19E--05 & 5.55E--04 & 1.31E--03 & 1.47E--10 & 1.09E--10 & 2.43E--03 & 8.73E--10 & 1.18E--08 & 9.62E--14 & 1.43E--13\\
 & & & & & & &1.88E--03 & 5.88E--03 & 8.74E--04 & 1.46E--08\\
$Si$ & 2.51E--02 & 3.41E--02 & 6.89E--02 & 2.79E--09 & 1.97E--09 & 1.66E--01 & 2.73E--09 & 3.48E--08 & 1.03E--09 & 1.05E--10\\
 & & & & & & &2.32E--01 & 1.24E--01 & 1.77E--03 & 2.89E--08\\
$S$  & 8.66E--03 & 1.35E--02 & 2.61E--02 & 1.18E--10 & 8.87E--11 & 1.09E--01 & 2.07E--09 & 6.36E--08 & 3.58E--13 & 4.22E--13\\
 & & & & & & &1.06E--01 & 4.59E--02 & 2.82E--06 & 4.32E--10\\
$Ca$ & 1.42E--03 & 3.01E--03 & 4.19E--03 & 8.57E--13 & 7.15E--13 & 2.38E--02 & 3.59E--10 & 5.76E--09 & 1.27E--26 & 1.17E--17\\
 & & & & & & &1.75E--02 & 8.78E--03 & 2.06E--09 & 5.71E--11\\
$Ti$ & 1.40E--04 & 1.71E--04 & 1.90E--04 & 3.27E--13 & 6.52E--14 & 3.25E--04 & 1.55E--10 & 3.40E--09 & 2.15E--26 & 1.33E--26\\
 & & & & & & &3.47E--04 & 4.80E--04 & 1.79E--09 & 1.42E--10\\
$Fe$ & 8.45E--02 & 2.00E--01 & 1.67E--01 & 4.93E--15 & 1.02E--17 & 1.74E--01 & 7.24E--11 & 1.32E--09 & 6.56E--36 & 1.16E--35\\
 & & & & & & &2.99E--01 & 3.35E--01 & 5.36E--10 & 7.01E--11\\
$^{56}Co$ & 8.15E--02 & 1.92E--01 & 1.62E--01 & 4.35E--20 & 3.03E--23 & 1.68E--01 & 2.14E--16 & 3.69E--15 & 8.19E--46 & 9.29E--46\\
 & & & & & & &2.91E--01 & 3.25E--01 & 1.54E--15 & 9.44E--16\\
\hline
\label{tabular:tab2}
\end{tabular}
\end{minipage}
\end{table*}
There the chemical composition of the ejecta is given at $2.5 \times 10^4$~s 
after the explosion, when  strong and electromagnetic reactions have ceased, 
but many nuclei have not yet decayed in their most stable form. Because dust 
formation occurs at $300 \div 600$ days after explosion, it is necessary
to take into account the radioactive decay of such nuclei. 
In the SN models of WW95 the energy of explosion can be adjusted to give 
the desired kinetic energy of the ejecta, typically $10^{51}$ erg. 
Following WW95, we explore the effects of 
$E_{kin}$ variation by considering a low ($E_{kin}=1.2 \times 10^{51}$~erg, Case
A) and a high ($E_{kin}=1.9 \times 10^{51}$~erg, Case B) value for this quantity.
We discuss separately the two cases in the following.

\subsection{Low Kinetic Energy (Case A)} 
\begin{figure}
\psfig{figure=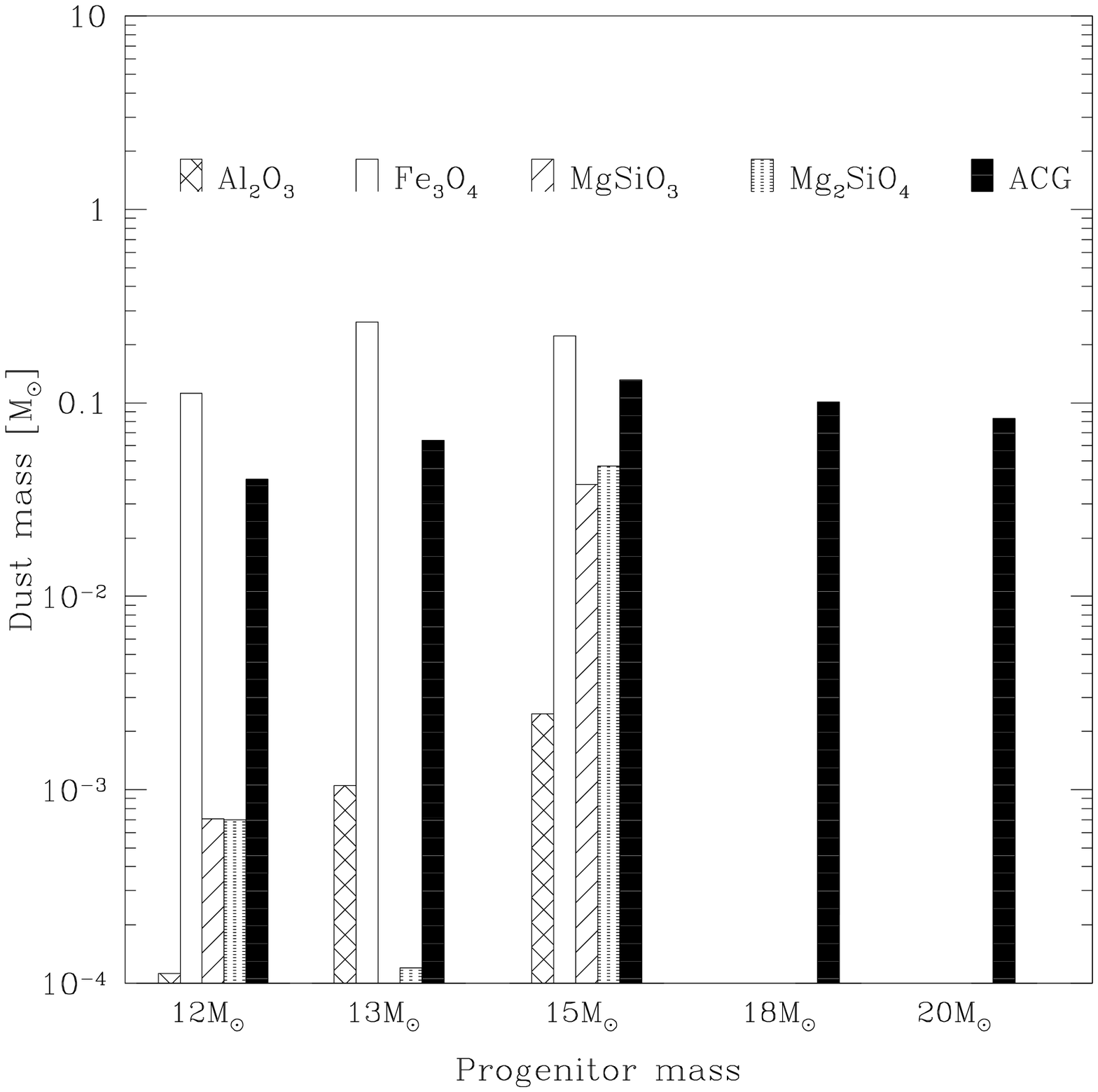,height=9cm}
\caption{Dust mass formed as a function of the SN mass (in the range $12 M_\odot < M < 20 
M_\odot$)        for initial
metallicity $Z=0$ and kinetic energy of the explosion $E_{kin}=1.2\times
10^{51}$~erg (Case A). Also shown is the grain composition.}
\label{fig3}
\end{figure}

\begin{figure}
\psfig{figure=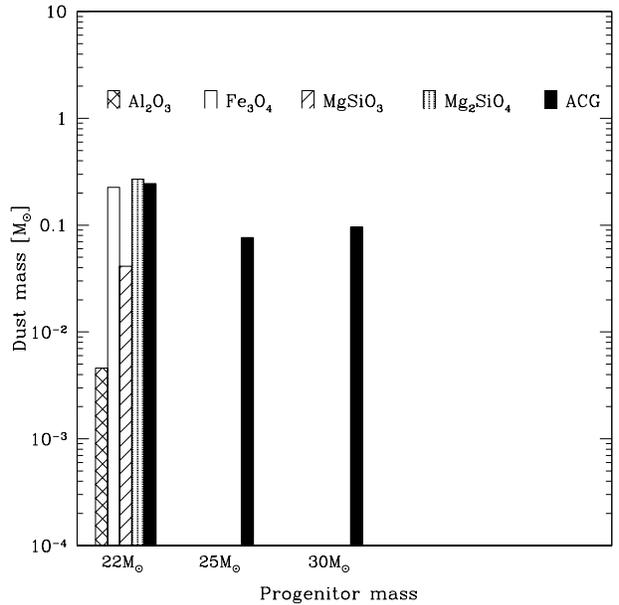,height=9cm}
\caption{Same as Fig. \ref{fig3}, but for the SN mass range $22 M_\odot < M < 30   
M_\odot$.}
\label{fig4}
\end{figure}

As the kinetic energy of the model is relatively low, this is not sufficient 
to completely expel the heavy elements external to the $Ni-Fe$ core of the
most massive SNe, and
a variable amount of material falls back onto the core, probably 
forming a neutron star or a black hole. The fallback will mostly affect
the inner layers, containing the heaviest elements; as a result, progenitors
with masses larger than $\approx 20 M_\odot$ will be prevented from
forming dust. For essentially the same reason, above $M \approx 15
M_\odot$, only AGC grains are formed. Figs. \ref{fig3}--\ref{fig4} show the amount of
dust formed as a function of progenitor mass, and the grain composition.
\begin{figure}
\psfig{figure=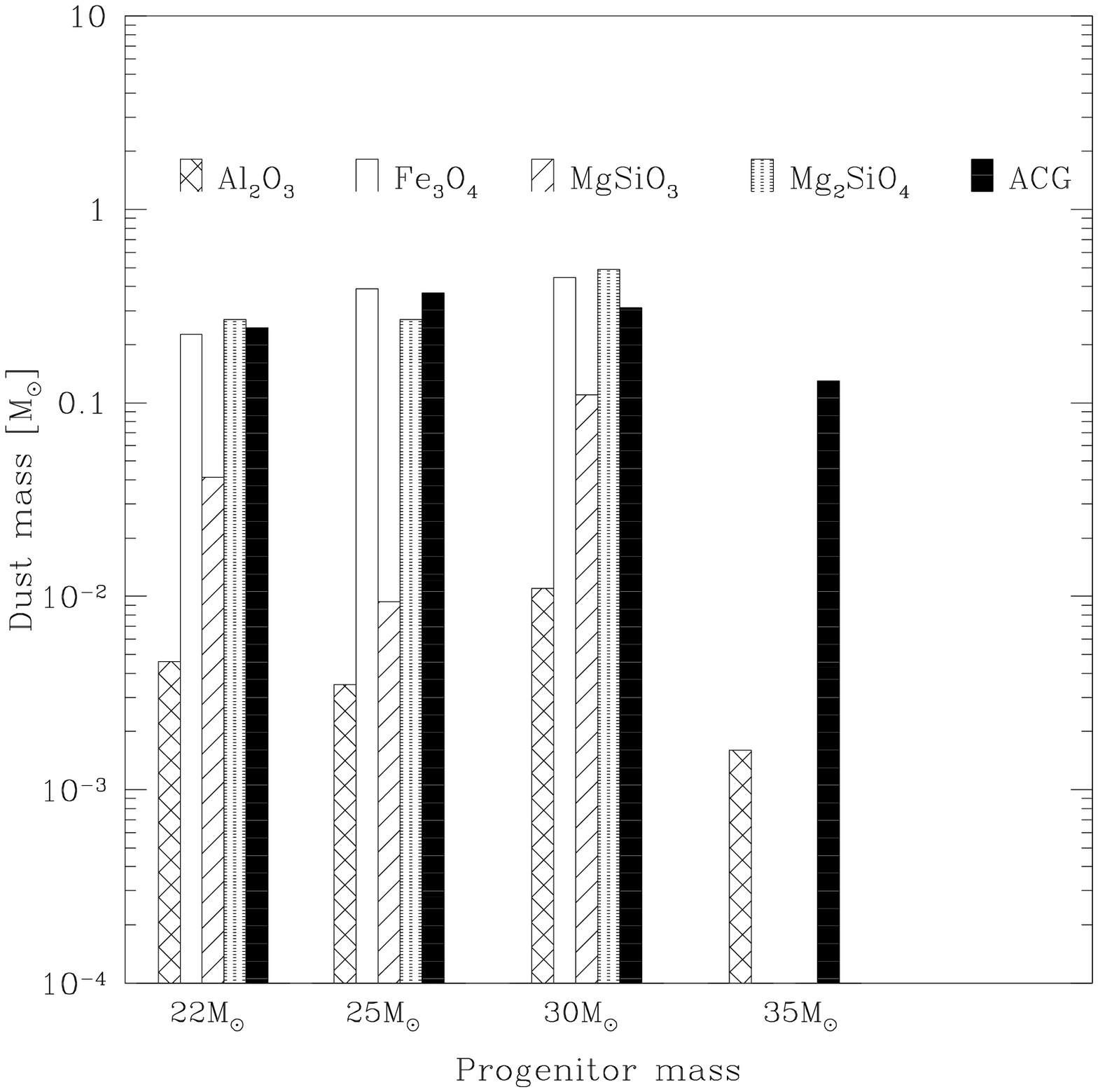,height=9cm}
\caption{Same as in Fig. \ref{fig3}, but for  kinetic energy of the explosion 
$E_{kin}=1.9\times 10^{51}$~erg (Case B);        the SN mass range is 
$22 M_\odot < M < 35  M_\odot$. For lower masses Case B gives the same results as Case A.}
\label{fig5}
\end{figure}
ACG are typically the first solid particles to condense, depending on the 
models.  The formation of these grains is quite fast with respect to the 
cooling time scale of the ejecta: most of the ACG dust mass forms in a narrow 
range of $30 \div 40$ K around $T = 1800$~K. 
Subsequently, at a temperature of $\approx 1600$~K
$Al_2O_3$ starts to condense, followed by $Fe_3O_4$, $MgSiO_3$ and $Mg_2SiO_4$ 
at $T \approx 1100$ K. Clearly this sequence is governed by the condensation
temperature of a given material. Also, the more refractory materials
preferentially end up into larger grains: this is because an  earlier formation
can exploit a higher concentration of the key species favoring the accretion 
process. The typical size of ACG dust grains is $a = 300$ \AA, whereas
$Fe_3O_4$ grains have typically $a= 20$ \AA \, and $Al_2O_3$,
$MgSiO_3$ and $Mg_2SiO_4$ grains are even smaller ($\approx 10$ \AA). 
In spite of the high condensation temperature,  $Al_2O_3$
grains do not grow to sizes comparable to those of ACG, as their growth is limited
by the low abundance of $Al$. In Figure \ref{fig6} grain sizes
of the most abundant compounds that form  in supernova ejecta are shown for 
four values of $Z$    (non zero metallicity cases are discussed later on).
The two silicates (enstatite and forsterite) start to condense almost simultaneously; however, $Mg_2SiO_4$ enters the supersaturation regime earlier than 
$MgSiO_3$. For this reason, forsterite grains grow quickly, 
strongly depleting the $Si$ (or $Mg$) available.
Figs. \ref{fig3}--\ref{fig4} clearly show that even starting from a primordial
composition, early SNe can         contribute a significant amount of
dust: for Case A, about $0.08 M_\odot \simlt M_d \simlt 0.3 \msun$ of dust/SN are
produced. Intermediate mass progenitors are the most efficient sources,
being able to convert up to 2\% of their mass into solid particles.
The reason is that they synthesize a considerable amount of heavy 
elements without suffering too much from the fallback process mentioned
above.

\subsection{High Kinetic Energy (Case B)} 
In this case the kinetic energy of the explosion for the progenitor
masses 25, 30, 35 and 40$\msun$ is chosen equal to $ 1.9 \times 
10^{51}$~erg; for lower masses the energy of explosion is the same as 
in case A. This energy is sufficient to
eject also the inner layers, which now can provide the elements to form
grains of various chemical composition, \ie not only ACG
as in case A. In Fig. \ref{fig5} we show the dust mass yield for Case B.
\begin{figure}
\psfig{figure=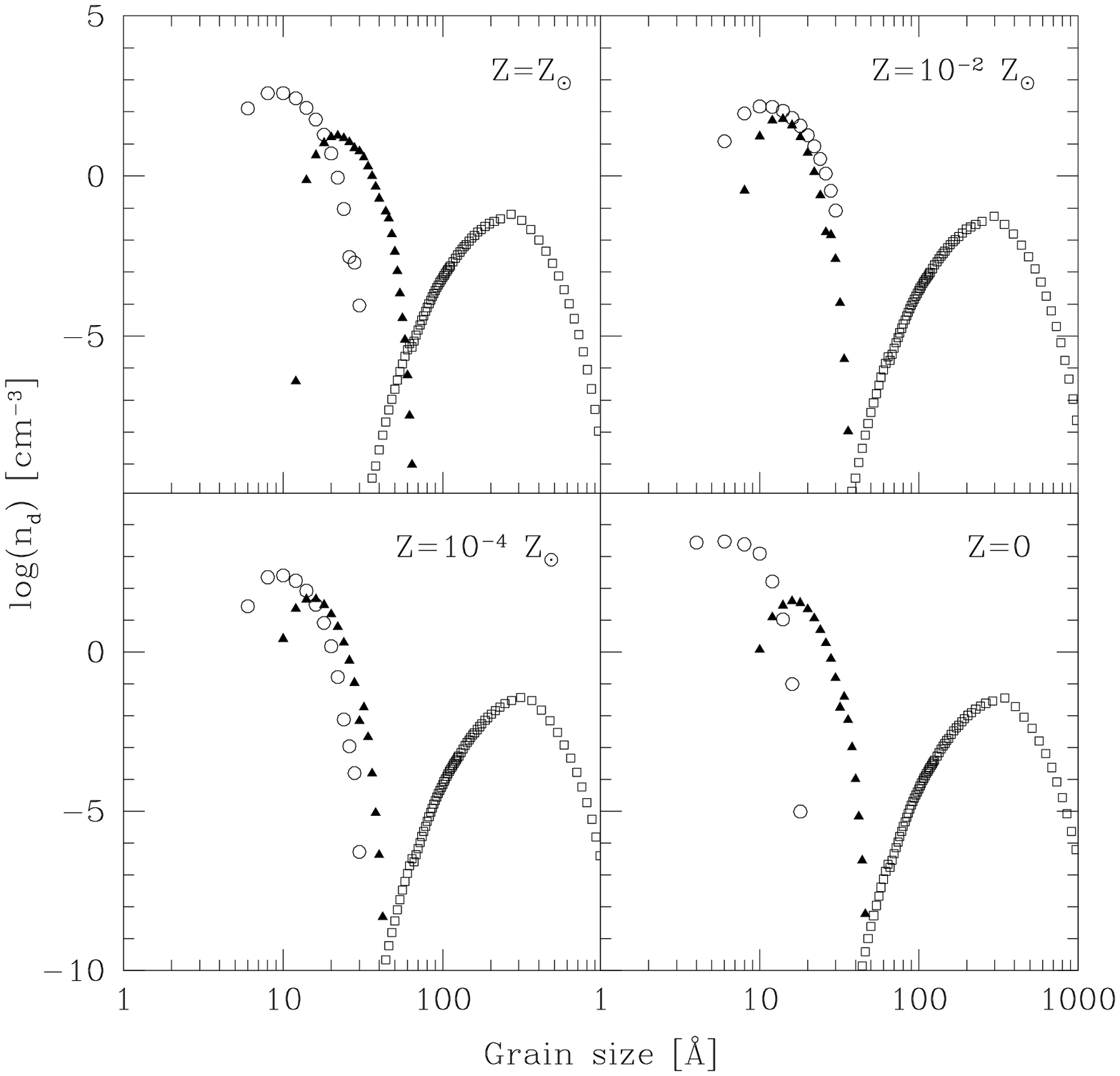,height=9cm}
\caption{Grain size distribution of ACG (open squares), $Fe_3O_4$ (solid triangles) and
$Mg_2SiO_4$ (open circles) grains for the $M=22\msun$ SN  model; results are give for four different metallicities of the progenitor.}
\label{fig6}
\end{figure}
Now SN up to masses $\approx 35 \msun$ are able to form dust. In
addition, a SN of $30 \msun$ is able to produce about $1.3 \msun$ of
dust (4.3\% of its mass). 
The formation sequence and the grain size distribution are 
very similar to those discussed for Case A. 

\subsection{Effects of $R_i$ and $\gamma$ Variations}
Up to now we have used the values of $R_i$ and $\gamma$ deduced from 
observations on SN 1987A. These values represent a reasonable
approximation but they might well depend on the specific properties
of the SN under examination.
\begin{figure}
\psfig{figure=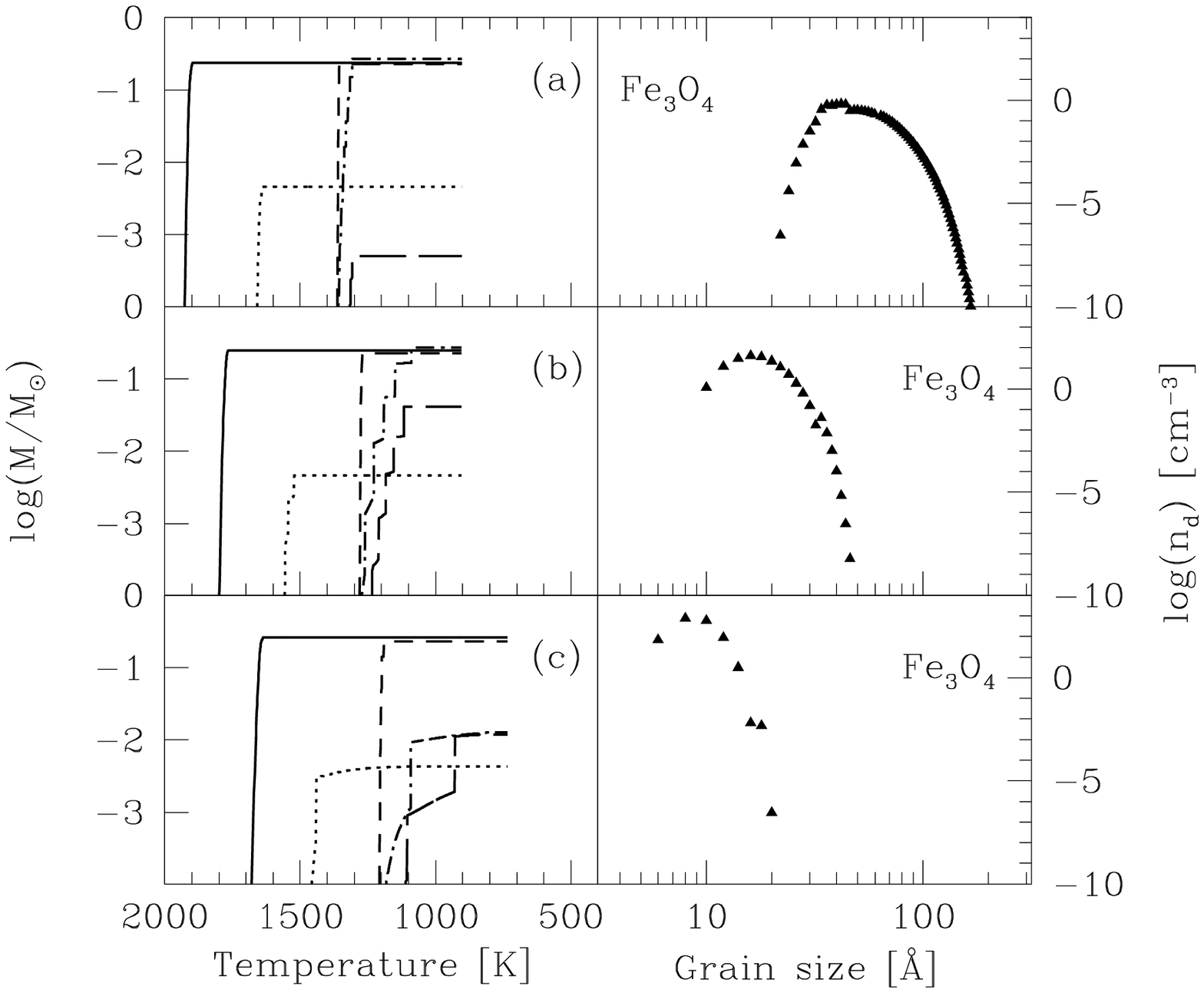,height=9cm}
\caption{Dust formation (left, line types as in Fig. \ref{fig1}) and $Fe_3O_4$ grain size distribution (right) for the $M=22\msun$ SN  model; the three cases refer to different values of $R_i$: 
(a) $7.5 \times 10^{14}$~cm, (b) $1.6 \times 10^{15}$~cm (standard value), 
(c) $3.4 \times 10^{15}$~cm. }
\label{fig7}
\end{figure}
Therefore, as a sanity check, we investigate the dependence of our results 
on different choices for these paramenters. As a benchmark, we focus  
on the  $M=22\msun$ SN model (Case A) and increase or decrease 
the standard value of $R_i=1.6\times 10^{15}$~cm by a factor 2.15; this  
corresponds to a variation of about 100 times in the initial volume of the ejecta. 
In Fig. \ref{fig7} we compare the dust formation evolution and the
final size of $Fe_3O_4$ grains for three values of $R_i$. The 
final masses of ACG, $Al_2O_3$ and $Fe_3O_4$ grains are almost unchanged
in the three cases because the gas density remains high enough 
for the collisional time scale (regulating the formation/accretion processes)
of  these materials to remain shorter than the expansion one. 
However, the behavior of silicates depends on the choice of $R_i$. 
As a general rule, $Mg_2SiO_4$ grains form first and grow faster 
than $MgSiO_3$ ones, thus using up the condensable materials
efficiently and ending up with a larger final total mass.
However, for larger values of $R_i$ (\ie larger volume, lower gas
density) this process is limited by the fact that the collisional scale
becomes longer, thus stopping the accretion at earlier times.
The grain size distribution shifts by about a factor 2  as $R_i$
is varied by a factor $(2.15)$. Thus the determination of  the 
grain size distribution is relatively uncertain.

The adiabatic index    $\gamma$ gives a measure of the ability of the gas
to cool: $\gamma$ greater than the standard value of 1.25 cause the gas 
to reach the supersaturation state when the volume of the ejecta is smaller. 
Therefore, a larger $\gamma$ case gives results similar to those
obtained for the low $R_i$ case discussed above. 
\begin{figure}
\psfig{figure=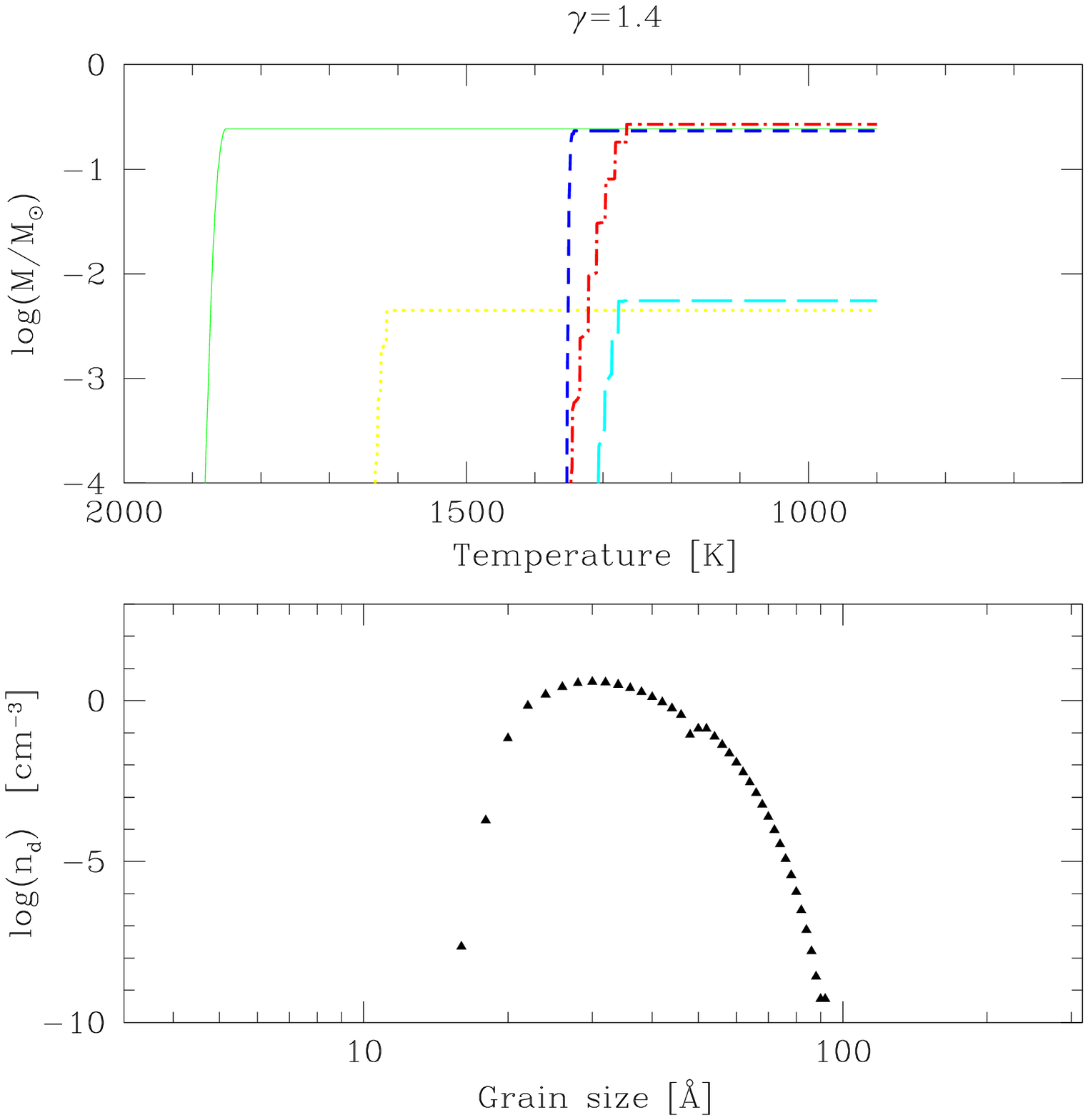,height=9cm}
\caption{Dust formation (top) and $Fe_3O_4$ grain size distribution (bottom)
for the $M=22\msun$ SN  with $\gamma=1.4$.}
\label{fig8}
\end{figure}
Figure \ref{fig8} 
shows the dust formation evolution and final grain size
distribution of $Fe_3O_4$ grains for the $M=22\msun$ model (Case A) with $\gamma=1.4$.
The similarity with the previous case with $R_i=7.4
\times 10^{14}$~cm and $\gamma=1.25$ is  evident. It has to be noted
though that the variation range for $\gamma$ (15\%) is smaller than that for $R_i$
(50\%), which might indicate that the dust formation process is more
sensitive to changes in the adiabatic index than in the initial radius.

\section{Extension to Higher  Metallicities}

We finally extend our results to non-primordial compositions by exploring the 
results for additional three metallicity values $Z/Z_\odot=10^{-4}, 10^{-2}, 1$.  
The chemical composition for these models is also  taken from WW95. 
We start by analyzing the dependence of the grain size distribution on 
metallicity. This is shown in the four panels of Fig. \ref{fig6}. 
Somewhat surprisingly, the dependence is almost absent, with grain radii ranging
from 5\AA~ to 0.1 $\mu$m for all values of $Z$. Also, the same material 
segregation is seen, with smaller grains being predominantly constituted by
silicate and magnetite and the larger ones made by amorphous carbon.   
This behavior can be explained by the fact that the final grain size is 
governed by the thermodynamics of the ejecta expansion (and therefore 
sensitive to $R_i$ and $\gamma$, as already pointed out before), but 
poorly affected by the ejecta composition. 

\begin{figure}
\psfig{figure=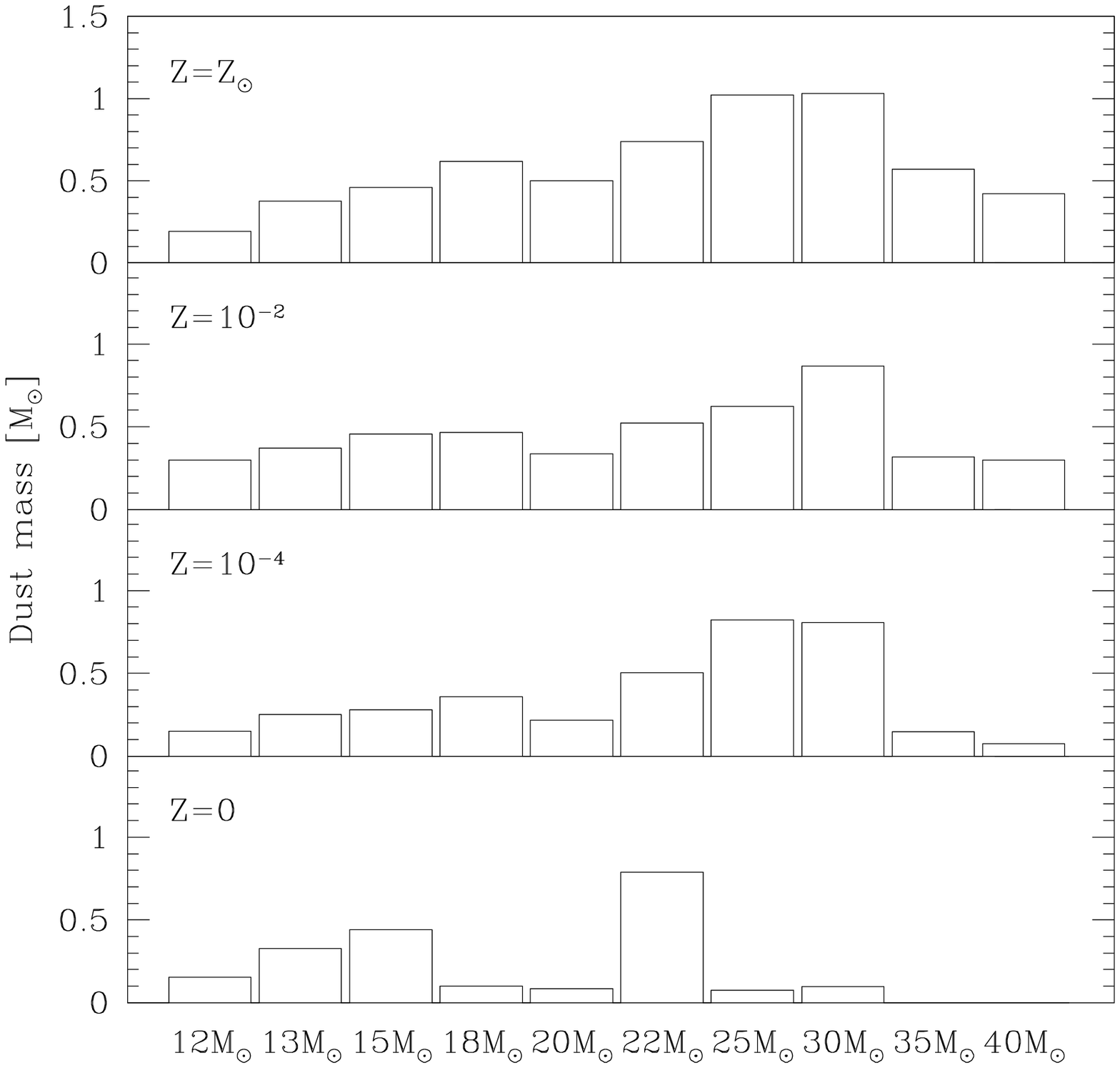,height=9cm}
\caption{Total dust mass produced as a function of SN mass and different metallicity 
of the progenitor (Case A).} 
\label{fig9}
\end{figure}

\begin{figure}
\psfig{figure=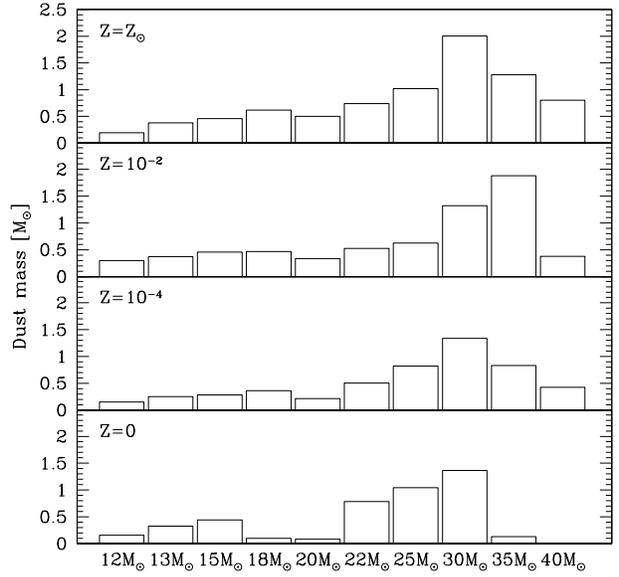,height=9cm}
\caption{Same of Figure \ref{fig9}, but for Case B.}
\label{fig10}
\end{figure}

The latter, instead, plays a more important role in determining the total 
amount of dust formed, as seen in Figs. \ref{fig9}--\ref{fig10} for 
Case A and Case B, respectively. Moving from $Z=0$ to higher metallicities
we observe that a large number of SNe contribute to dust production: a clear 
example of this is the behavior of SNe with mass 18-20 $M_\odot$, which
increase their dust yield from $< 0.1 M_\odot$ up to $0.5-0.6 M_\odot$ for $Z=1$.  
The enhancement of dust formation is due to the fact that the density of heavy 
elements in the ejecta becomes large enough to allow the state of supersaturation
to be reached more easily. 
This trend results in a steady increase of the total amount of the dust produced  
in the four cases (obviously, when appyling these results one has to weigh over 
the appropriate IMF); they are (2.06, 3.6, 4.5, 5.9) $M_\odot$ for $Z/Z_\odot=
(0,10^{-4}, 10^{-2}, 1)$, respectively,  for Case A.
However, it is interesting to note that at each metallicity, the maximum amount of
dust produced by a single SN varies little, and it is never higher than 
approximately $1 M_\odot$. Finally, the differences between Casa A and Case B are
found to be minor.

\section{Summary and Discussion}                     

We have investigated the formation of dust in Type II supernovae mostly with
primordial abundances, a property  characterizing these events
in the early universe; however, we have also considered non-zero 
metallicity values up to $Z=Z_\odot$. The calculations are based on standard
nucleation theory and the scheme has been first tested on the well studied case
of SN1987A, yielding results that are in satisfactory agreement with the available
data (see Section 3). The main results of the paper are the following:

\begin{itemize}

\item The first solid particles in the universe are formed by Type II SNe. 
The dust grains are made of silicates (predominantly $Mg_2SiO_4$),
amorphous carbon (ACG), magnetite ($Fe_3O_4$), and corundum ($Al_2O_3$) and form
about 300-660 days after explosion.

\item The largest grains are the ACG, with sizes around 300~\AA~, whereas other
grain types have smaller radii, around 10-20~\AA. The grain size distribution 
depends considerably on the thermodynamics of the ejecta expansion (characterized by
their initial radius and adiabatic index) and variations in the results by a
factor $\approx 2$ might occur within the estimated range of $R_i$ and $\gamma$. 
Also, and for the same reason, the grain size distribution, is essentially
unaffected by metallicity changes.  

\item The amount of dust formed is instead very robust to variations in $R_i$ 
and $\gamma$. For $Z=0$, we find that SN with masses in the range (12-35)$M_\odot$
produce about $0.08 \msun \simlt M_d \simlt 0.3 M_\odot$of dust/SN in the low kinetic energy 
explosion case; slightly higher final yields are obtained in the high kinetic
energy case. The above range increases by roughly 3 times as the metallicity
is increased to solar values. 
\end{itemize}

The previous results clearly show that it is likely that dust has been
present in the universe immediately after the first stars appeared. 
This has a large number of consequences that it will be necessary to
study in detail. Among possible effects, the most outstanding ones concern
the opacity of the universe at high $z$, spectral distorsions in the 
Cosmic Microwave Radiation caused by dust re-emission of absorbed UV-optical
light, catalyzation  of H$_2$ molecular hydrogen formation  and heavy elements
depletion in the interstellar medium of pristine galaxies and in the 
intergalactic medium.

The first two issued were already discussed in detail by Loeb \& Haiman (1997) 
and Ferrara \etal (1999), and we defer the interested readers to those works for
details. In short, the expected IGM opacity contributed by dust 
around the observed wavelength $\lambda \sim 1 \mu$m is $\sim 0.13$ and 
it rapidly increases to $\approx 0.35$ at $z=20$.
The expected CMB spectral distortions due to high-$z$ dust
is only $\sim 1.25-10$ times smaller than the current COBE upper limit, but these 
numbers might depend crucially on the formation epoch and abundance of dust.  

\begin{figure}
\psfig{figure=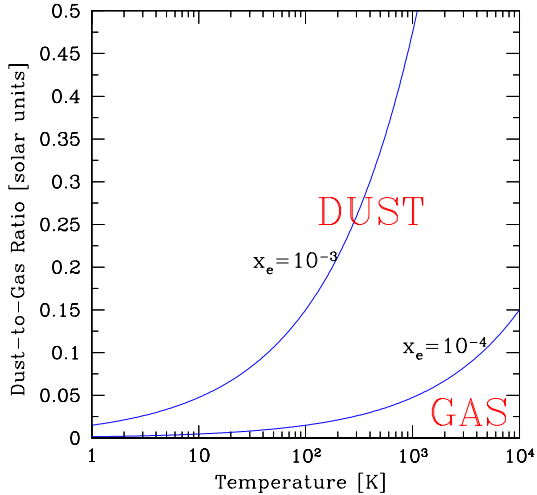,height=9cm}
\caption{Comparison between molecular hydrogen formation rates in the gas phase
and on dust grain surfaces as a function of the dust-to-gas ratio and gas 
temperature. Above the two curves, indicating the equality between the rates 
for two different values of the electron fraction $x_e=10^{-3}-10^{-4}$, 
H$_2$ formation on grains dominates.}
\end{figure}

In addition to the above effects, Type II SNe can also initiate molecular 
hydrogen formation on dust grain surfaces rather than in the gas phase,
the second process being the only viable in a dust free environment. 
As already mentioned, at high redshift Type II SNe are 
the only possible sources of dust, due to the short age of the universe 
and the long evolutionary timescales characterizing more conventional dust 
sources, as for example evolved  stars.
Thanks to the above results we can now answer the following
question: what is the minimum amount of dust required in order for the
molecular hydrogen formation on grains to become competitive with the
gas phase one ? An order-of-magnitude answer can be obtained by comparing the
two formation rates. At the low densities relevant   here,
H$_2$ is formed in the gas phase mainly
via the channel $H+ e^- \rightarrow H^- +h\nu$, at rate $k_8$
(the rate coefficient $k_8$ is given in Abel \etal 1997);
formation via the H$_2^+$ channel, when
included, is found to be negligible in our case. Therefore the formation rate
in the gas phase is ${\cal R} \simeq k_8 n_{H^-} n_H$. The formation rate on
grain surfaces is instead given by ${\cal R}_d \simeq 0.5 \langle \gamma
c_s \sigma\rangle n_d n_H$, where $\gamma$ is the sticking coefficient,
$c_s$ is the sound speed in the gas, and $\sigma$ is the grain cross section.
The equality between the two rates can be cast into the following form:
${\cal D} = 0.1 \sqrt{T} x_e$, where ${\cal D}$ is the dust-to-gas ratio
normalized
to its Galactic value, $T$ is the gas temperature and $x_e$
the gas ionization fraction. 
 For typical parameters of the PopIII objects (Ciardi \etal 2000),
H$_2$  production on dust grains becomes dominant once ${\cal D}$ is larger than
5\% of the local value. With the dust yields calculated above, we then
conclude that only about 50 SN are required to enrich in dust to this level
a primordial object. 
Clearly, early dust formation might play a role in
the formation of the first generation of objects.

Even in larger galaxies, which will form later on when the overall
metallicity and dust levels in the universe have increased, dust will 
be at least as important. For example, on the scale of the molecular
clouds in a galaxy, it will provide the opacity to stop the infall on
forming protostars, hence possibly changing the properties of the IMF,   
and to allow the cloud to self-shield from damaging H$_2$ photodissociating
UV radiation. Finally, dust photoelectric heating is known to be the major
heating source for the diffuse ISM, and hence partecipating to the onset
of its observed multiphase structure (Ricotti, Ferrara \& Miniati
1997, Spaans \& Norman 1997).   

The fate of the dust that we predict from Type II SNe has yet to be determined.
What fraction of the grains will be able to survive the passage through the
reverse shocks at which the ejecta are thermalized ? 
Grains in a hot gas are essentially destroyed via thermal sputtering, \ie 
collisions with ions or electrons with Maxwellian velocity distribution.
However, in spite of the still poorly understood underlying physics, it seems
unlikely that the efficiency of grain destruction in an adiabatic shock can be higher
that 10\% (McKee 1989). Grains are more likely to be destroyed behind radiative 
shocks, by the combined effects of a greatly enhanced gas density and betatron
acceleration that increases the grain Larmor frequency. However, if, as expected,
the magnetic field is weak at high $z$, the efficiency cannot be very high.
The grains then will follow the fate of the gas and will likely be expelled in the IGM, as PopIII object suffer complete blowaway of their gas (Ciardi \etal 2000).
Once in the intergalactic space, dust might have strong influence, for example,
on the determination of cosmological parameters via the observation of 
high $z$ (Type I) SN (Aguirre 1999,  Croft \etal 2000).

As a final remark, we have seen that SNe with mass
above $M=15 M_\odot$ predominantly form amorphous carbon grains (the ejecta of these stars have higher C/O ratios); in doing so,
they use up virtually all the available carbon yield in the ejecta. This implies
that carbon in the IGM at high redshift will be strongly depleted above redshifts
at which only Type II SNe contribute to the metal enrichment of the universe.
It will be then posible to test this prediction once a sample of target sources
at $z \simgt 5$ will become available for absorption line studies.

\vskip 0.5truecm
This work was completed as one of us (AF) was a Visiting Professor at
the Center for Computational Physics, Tsukuba University, whose support
is gratefully acknowledged.

\end{document}